\documentstyle[12pt,epsfig]{article}
\input{epsfig.sty}

\begin{document}

\begin{titlepage}

\begin{flushright}
UB-ECM-FP-79/12\\
ICCUB-12-324
\end{flushright}

\vspace*{2cm}

\begin{center} 
{\Large {\bf Feshbach resonance: a one dimensional example.}}\\[1.5cm]
\large {{\bf Josep Taron }}\footnote{e-mail: taron@ecm.ub.es}\\[1cm] 
Departament d'Estructura i Constituents de la Mat\`{e}ria\\
Facultat de F\'{\i}sica, Universitat de Barcelona\\
and\\
Institut de Ci\`encies del Cosmos\\[0.5cm]
Diagonal 645, E-08028 Barcelona, Spain.
\end{center}

\vspace*{1.0cm}
\begin{abstract}
We present a simple one-dimensional example of a spin 1/2 particle
submitted to a delta-type potential which interacts differently with the two
components of the wavefunction and to an external magnetic field.
It has
two coupled channels,
admits a closed solution and features the Feshbach resonance
phehomenon by proper tunning of the magnetic field.
\end{abstract}

\vspace*{3.0cm}


\vspace*{3cm}

\end{titlepage}


\section{Introduction}
Consider a spin 1/2 uncharged particle with its motion confined
to the $z$-axis only, driven by the hamiltoninan:
\begin{equation}
H = - {\rm{\bf I}} \; \frac{\hbar^2}{2m} \frac{d^2}{dz^2}+
               \left( \begin{array}{cc}
                      V_1(z) & 0 \\ \nonumber
                       0 & V_2(z) 
                \end{array}  \right) -\mu B \sigma_x.
\label{hamiltonian}
\end{equation}
It consists of a kinetic term, a potential term that we assume is short range
which interacts differently with the spin up and the spin down components
of the wavefuntion and the interaction with a constant external magnetic
field that points in the $x$-direction $\vec{B}= (B,0,0)$; 
$\mu$ is the particle dipole magnetic moment,
$\sigma_x$ is the $x$ Pauli matrix and $I$ is the $2 \times 2$ identity 
matrix.

The time independent Sch\"odinger equation for such a system reads,
in terms of the spin up $\psi(z)$ and the spin down $\varphi(z)$
components of the wavefunction,
$$-\frac{\hbar^2}{2m}\frac{d^2 \psi(z)}{dz^2} + V_1(z) \; \psi(z) 
-\mu B \; \varphi(z)= E \; \psi(z),$$
\begin{equation}
-\frac{\hbar^2}{2m}\frac{d^2 \varphi(z)}{dz^2} + V_2(z) \; \varphi(z) 
-\mu B \; \psi(z)= E \; \varphi(z),
\label{sch}
\end{equation}
where $E$ is the total energy of the system.
By adding and subtracting these equations
one finds,
$$-\frac{\hbar^2}{2m}\frac{d^2 u(z)}{dz^2} + V \; u(z)+
W \; v(z)= (E+\mu B) \; u(z),$$
\begin{equation}
-\frac{\hbar^2}{2m}\frac{d^2 v(z)}{dz^2} +  V \; v(z)+
W\; u(z)= (E-\mu B) \; v(z),
\label{sch2}
\end{equation}
with,
\begin{equation}
V \equiv \frac{V_1+V_2}{2}, \;\;\; W\equiv \frac{V_1-V_2}{2},
\label{def}
\end{equation}
where the
$u=(\psi+\varphi)/\sqrt{2}$ and $v=(\psi-\varphi)/\sqrt{2}$ combinations 
are the components in the basis of eigenvectors of $\sigma_x$, which 
diagonalize the magnetic part of the interaction and define
the spin Zeeman states of the particle, with
magnetic energies $\pm \mu B$.  

The different {\it channels}
are defined by specifying these Zeeman states.
A channel is said to be {\it open}
or {\it closed} depending on the sign of the combination
$E\pm \mu B$ in the r.h.s of (\ref{sch2}), positive or negative,
respectively. 
As we shall see, it is crucial that the magnetic energy of the two Zeeman 
components have opposite signs. This asymetry allows the possibility of 
having simultaneously one open and one closed channel.

We can think of the coupling interaction term $W$ as a
perturbation that modifies the two {\it bare} decoupled equations, 
one for each of
the components. It is said that the {\it bare} equations are {\it dressed}
by the coupling $W$.

In a scattering experiment, with an incident beam of pure $u$ component
in a remote zone far from the region where the potential acts,
the $u$ channel is open
and its incident kinetic energy $Q$ corresponds to the combination
$Q= E+\mu B$, which can be fixed experimentaly. This in turn fixes
the r.h.s. of the second equation too, $E-\mu B=Q-2 \mu B$, whose sign can be
changed by conveniently tunning the magnitude of the external field $B$.

Finally, notice that the eqs.
(\ref{sch2}) do not decouple unless $V_1=V_2$. 

\section{Delta-type potentials.}

We consider {\it attractive} delta-type potentials
$V_1(z)= - g_1 \delta(z)$ and $V_2(z)=-g_2 \delta(z)$, with $g_1 \neq g_2$,
where both $g_1$ and $g_2$ are taken positive. 
One can think of it as a mathematical effective substitute that describes the
low energy scattering off a pottential which only acts 
in a small region, much smaller than the wavelength of the scattered 
particle \cite{lipkin}.

In terms of the combinations,
\begin{equation}
\alpha_1 =\frac{m g_1}{\hbar^2}, \;\;\;
\alpha_2 =\frac{m g_2}{\hbar^2}, \;\; {\rm and} \;\;
K^2= \frac{2 m (E+\mu B)}{\hbar^2},\;\;\;\; 
K'^2= \frac{2 m (E-\mu B)}{\hbar^2},
\label{Ks}
\end{equation}
equations in (\ref{sch2}) can be written as,
\begin{equation}
\left( \frac{d^2}{dz^2}+ K^2 \right) \; u(z)=-S \; \delta(z), \;\;\;
\left( \frac{d^2}{dz^2}+ K'^2\right) \; v(z)=-S' \; \delta(z),
\label{sch3}
\end{equation}
where, 
\begin{equation}
S=(\alpha_1 + \alpha_2) \; u(0)+ (\alpha_1 - \alpha_2) \; v(0),\;\;
\;\;
S'=(\alpha_1 + \alpha_2) \; v(0)+ (\alpha_1 - \alpha_2) \; u(0);
\label{Ss}
\end{equation}
beeing $u(0)$, $v(0)$ the values of the wavefunction components at $z=0$, 
the only place where the potential acts.

There are various possibilities that combine different signs of $K^2$ and 
$K'^2$.
Of special interest to us is the case with one open channel, coupled
to the other channel closed; the first has $K^2>0$, 
whereas the second has $K'^2 <0$.
We set $K'=i \beta'$, with $\beta'$ real and positive.
The scattering solution, with the usual 
incoming wave $\displaystyle e^{iKz}$ 
at the open channel entrance plus
outgoing scattered waves, may be written as
(see Appendix),
\begin{equation}
u(z)=e^{iKz}-S \; \frac{e^{iK|z|}}{2iK}=
e^{iKz}-\frac{1}{iK}\left[  \frac{\alpha_1 + \alpha_2}{2} \; u(0)+ 
\frac{\alpha_1 -\alpha_2}{2}\; v(0) \right] \; e^{iK|z|},
\label{sol1u}
\end{equation}
and for the closed channel,
\begin{equation}
v(z)=S' \frac{e^{-\beta'|z|}}{2\beta'}=\frac{1}{\beta'}
\left[ \frac{\alpha_1 + \alpha_2}{2} \; v(0)+ \frac{\alpha_1 - \alpha_2}{2} \; u(0)\right]\; e^{-\beta' |z|}.
\label{sol1}
\end{equation}
The {\it sources} in the r.h.s. of (\ref{sol1}) are proportional
to the values that the wavefunction components $u(0)$, $v(0)$ take at the
origin, which can be obtained self-consistently from (\ref{sol1})
by setting $z=0$ and solving the resulting linear system.
We obtain,
\begin{equation}
v(0)=\frac{1}{2}\;\frac{\alpha_1-\alpha_2}{\displaystyle \beta' - \frac{\alpha_1+\alpha_2}{2}} \; 
u(0),
\label{v0}
\end{equation}
and from the first equation in (\ref{sol1}),
\begin{equation}
u(z)= e^{iKz}- \alpha_{\rm{eff}}(\beta') u(0) \; \frac{e^{iK|z|}}{iK}.
\label{u} 
\end{equation}
where we have defined,
\begin{equation}
\alpha_{\rm{eff}}(\beta')= \frac{\alpha_1+\alpha_2}{2}+\frac{1}{4}\;
\frac{\displaystyle\left(\alpha_1-\alpha_2\right)^2}{\displaystyle \beta'- \frac{\alpha_1+\alpha_2}{2}}.
\label{alpha.eq}
\end{equation}

\begin{figure}[ht]
\begin{center}
\mbox{\epsfig{file=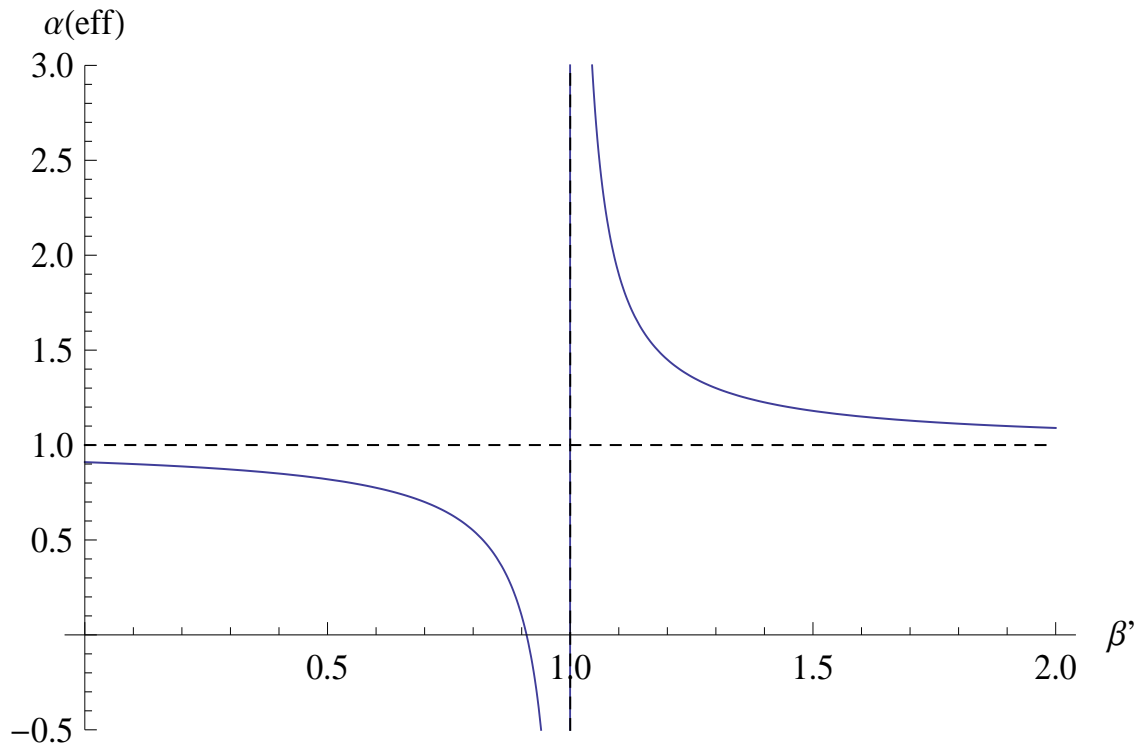,height=10cm,width=15cm,clip=,angle=0,silent=}}
\end{center}
\caption{Plot of $\displaystyle \alpha_{{\rm eff}}(\beta')$ in 
(\ref{alpha.eq}). For fixed $K$, by varying the magnetic field $B$
and thus $0\leq \beta' < \infty$, 
$\displaystyle \alpha_{{\rm eff}}$ can take on any
value in the full range, $-\infty < \alpha_{{\rm eff}}< + \infty$.
Notice the pole
at $\beta_{{\rm pole}}'$, and the value of $\beta_c'$ at which 
$\displaystyle \alpha_{{\rm eff}}$ vanishes. [The plot corresponds
to the values $\alpha_1=1.3$, $\alpha_2=0.7$ in units of inverse length
$\displaystyle \left[(\alpha_1+\alpha_2)/2\right]$ - which corresponds to the
inverse of the spatial extension of the wavefunction closed channel 
component (\ref{sol4}) at the Feshbach resonance. We have $\beta'_c=0.91$
and $\beta_{{\rm pole}}=1$ in such units. ]}
\end{figure}


Equations (\ref{u}) and (\ref{alpha.eq}) summarize the remarkable 
result that we wish to emphasize: the net result of the two coupled channels, 
with one channel open and the other one closed, 
is equivalent to a single channel
one-dimensional scattering problem with wavenumber $K$ and an effective
delta-type potential given by (see Appendix),
\begin{equation}
V_{{\rm eff}}(z)=-g_{{\rm eff}}(\beta') \;\delta(z), \;\;  
{\rm with \;\; coupling}\;\;
g_{\rm{eff}}(\beta')\equiv\frac{\hbar^2}{m}\alpha_{\rm{eff}}(\beta'),
\label{Veq}
\end{equation} 
that depends on 
$\displaystyle \beta'(K,B)=\sqrt{\frac{2m}{\hbar^2} (2\mu B)- K^2}$. 
Therefore, for fixed $K$ any value in the whole range,
$-\infty < \alpha_{\rm{eff}}(\beta') < +\infty$, is available for the
effective coupling by varying $B$ \break (see Fig. 1).
This effect disappears completely if
$\alpha_1 = \alpha_2$, in which case the equations in (\ref{sch2}) are not
coupled anymore.

We find,
\begin{equation}
u(0)=\frac{1}{1+{\displaystyle \frac{\alpha_{{\rm eff}}(\beta')}{iK}}},
\label{uv}
\end{equation}
and the complete solution (\ref{sol1}) reads,
\begin{equation}
u(z)=e^{iKz}+r(K,\beta') e^{iK|z|}, \;\;\; {\rm with} \;\;
r(K,\beta')=
-\frac{\alpha_{{\rm eff}}(\beta')}{iK+\alpha_{{\rm eff}}(\beta')},
\label{sol2}
\end{equation}
where $r(K,\beta')$ and 
$t(K,\beta')=1+r(K,\beta')$ are the reflection and the transmission
amplitudes, respectively. For the other component,
\begin{equation}
v(z)=v(0) e^{-\beta'|z|}=\frac{iK \left(\displaystyle{\frac{\alpha_1-\alpha_2}{2}}\right)}{\left(iK + \displaystyle{\frac{\alpha_1+\alpha_2}{2}}\right)\left(\beta'-\displaystyle{\frac{\alpha_1+\alpha_2}{2}}\right)+\left({\displaystyle{\frac{\alpha_1-\alpha_2}{2}}}\right)^2}\; e^{-\beta'|z|}.
\label{v}
\end{equation}
\vspace*{0.5cm}
There is a value $\beta'_c$ 
which makes the effective coupling vanish,
$\alpha_{\rm{eff}}(\beta_c')=0$,
\begin{equation}
\frac{1}{\beta_c'}= \frac{1}{2} \left( \frac{1}{\alpha_1} + \frac{1}{\alpha_2}
\right),
\label{beta.c}
\end{equation}
where the net effective interaction in the open channel disappears 
and $r(K,\beta'_c)=0$. 
The wave function becomes,
\begin{equation}
u(z)=e^{iKz},\;\;
v(z)=-\frac{\alpha_1+\alpha_2}{\alpha_1-\alpha_2} \; e^{-\beta'_c|z|},
\label{sol3}
\end{equation}
with a persisting presence in the closed channel in spite of 
$\alpha_{{\rm eff}}(\beta'_c)$ being zero.

Moreover, $\alpha_{\rm{eff}}(\beta')$ has a pole at,
\begin{equation}
\beta_{{\rm pole}}'=\frac{(\alpha_1+\alpha_2)}{2}, 
\label{pole}
\end{equation}
where it diverges, which leads to 
total reflection for any value of $K$, $r(K,\beta'_{{\rm pole}})=1$, 
with no transmission whatsoever. This is the Feshbach resonant
solution. Its wavefunction is,
\begin{equation}
u(z)=e^{iKz}- e^{iK|z|},
\label{sol4u}
\end{equation}
i.e.,
\begin{equation}
u(z<0)= 2 i \sin Kz; \;\; u(z>0)=0,
\label{sol4ubis}
\end{equation}
and,
\begin{equation}
v(z)=\frac{2iK}{\alpha_1-\alpha_2} \; e^{-\beta'_{{\rm pole}}|z|}.
\label{sol4}
\end{equation}
Notice that
$\beta'_{{\rm pole}}$ 
coincides precisely with the bound state value (see (\ref{quantization}) 
in the Appendix) in  the {\it bare} closed
$v$-channel.

It is worth stressing once more 
that across these two values of $\beta'$, $\beta'_c$
and $\beta'_{{\rm pole}}$,
the effective coupling $\alpha_{{\rm eff}}(\beta')$ changes sign,
i.e., the character
of the interaction changes from attractive to repulsive.
\vspace*{0.5cm}

The reflection coefficient (which in one dimension plays an analogous
role to that of the cross section in three dimensions),
\begin{equation}
|r(K,\beta')|^2=\frac{1}{1+(K/\alpha_{{\rm eff}}(\beta'))^2},
\label{refl}
\end{equation}
has its peak at $K=0$ with a width equal to $|\alpha_{{\rm eff}}|$,
and is insensitive to the sign of $\alpha_{{\rm eff}}$. As we approach
the Feshbach point $\beta'_{\rm pole}$, 
$\alpha_{{\rm eff}}$ diverges, the reflection coefficient broadens
up, and it becomes flat and equal to unity at the 
pole\footnote{In the literature 
of cold atomic gases, this is  
referred to as the {\it unitarity point}.}.

Let us complete the discussion with a comment on the
partial wave phase shifts. In one dimension
and for potentials that are even functions of $z$, two phase shifts
$\delta_0$, $\delta_1$ encode all the information concerning scattering, one
for each sector of
even and odd functions, respectively.
They relate to the reflection and transmission amplitudes by the expressions
\cite{lipkin},
\begin{equation}
r=\frac{1}{2} \left( e^{2 i \delta_0} - e^{2 i \delta_1} \right), \;\;
t=\frac{1}{2} \left( e^{2 i \delta_0} + e^{2 i \delta_1} \right).
\label{phase.shifts}
\end{equation}
From (\ref{sol2}) we find,
\begin{equation}
\cot \delta_0= K/\alpha_{{\rm eff}},
\label{cot}
\end{equation}
and $\delta_1=0$ \footnote{The $\delta_1$ shift is associated to the odd part
of the wave function. In this case, the $\sin Kz$ part of the incoming
wave in (\ref{u}) vanishes at $z=0$, which is the only place it feels
the delta-type potential. Therefore, the odd part does not suffer any
interaction and its corresponding phase shift $\delta_1$ vanishes.}.
At the Feshbach resonance point we find $\cot \delta_0=0$, i.e., 
$\delta_0=\pi/2$. At $\beta'_c$, $\cot \delta_0 \to \infty$, i.e.,
$\delta_0=0$, according to the noninteracting situation.

\vspace*{0.5cm}

For the value of $\beta'_{{\rm pole}}$ our example features 
the so called Feshbach resonance effect
\cite{F2}, \cite{F3} (see \cite{CTGO};  \cite{DS}, \cite{KGJ}, \cite{T}, 
\cite{HS}, 
\cite{CGJT} for some recent reviews).
The incoming state of the particle in the open $u$-channel is coupled by
the interaction $W$ in (\ref{sch2}) to the bound state $v_b$, hold
in the {\it bare} $v$-closed channel.
This colliding $u$-beam
can thus make a virtual transition to $v_b$,
the duration of which scales as 
$\hbar/\Delta E$, as the inverse of the detunning $\Delta E$, i.e., the 
difference 
between the incident energy in (\ref{sch2})
$E+\mu B$ and the magnetically shifted 
energy of the bound state $E_b-\mu B$. When 
the magnetic field is such that the denominator  $\Delta E$ 
is close to zero, the virtual
transition can last a very long time and this enhances the scattering
amplitude \cite{CTcourse}.

Therefore, when the channels are coupled the total scattering amplitude can
be viewed as the sum of
a direct one, of the incident $u$-beam scattering off  
the potential $V$ in (\ref{sch2}), and an indirect
one just explained above, due to the possibility of 
virtually flippling the
$u$ component to $v$, provided by the coupling $W$.
The amplitudes interfere and give rise to an effect which is 
constructive and 
enormously enhanced when the magnetic field is such that 
$\beta'=\beta'_{{\rm pole}}$ (Feshbach resonance) 
and destructive for the value at $\beta'=\beta_c'$.

\vspace*{0.5cm}


\begin{figure}[ht]
\begin{center}
\mbox{\epsfig{file=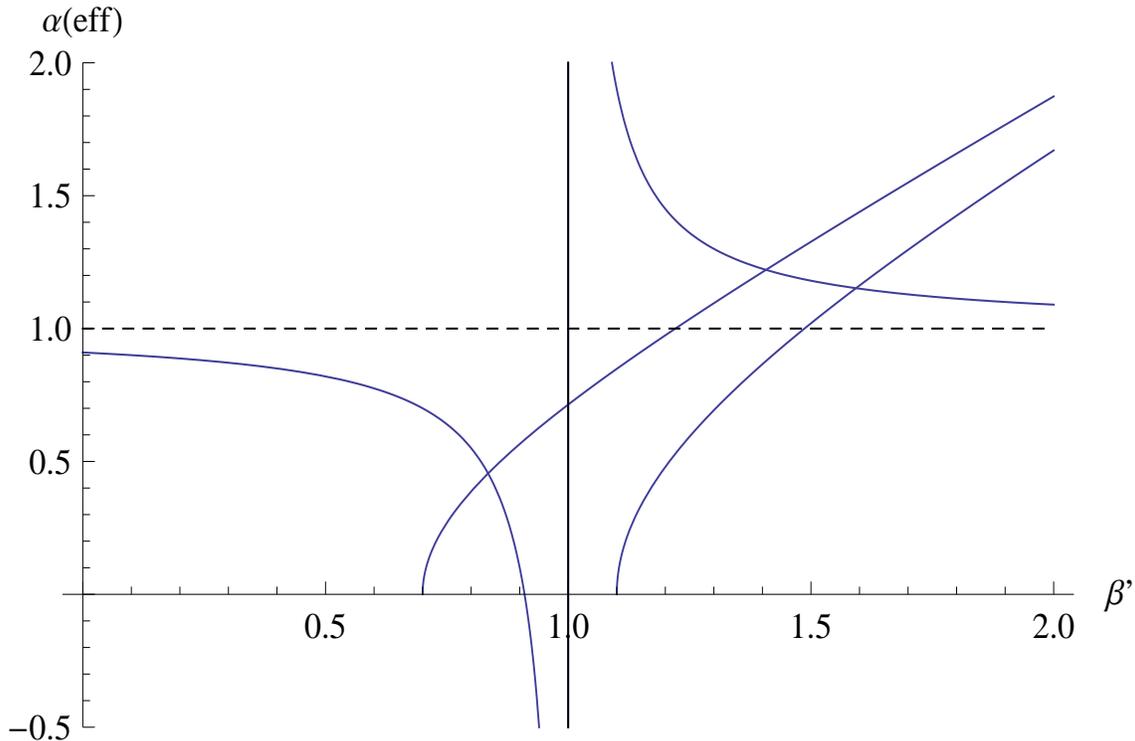,height=10cm,width=15cm,clip=,angle=0,silent=}}
\end{center}
\caption{Graphical solution of (see Eq.(\ref{bound})):
$\displaystyle \alpha_{{\rm eff}}(\beta')=\beta(\beta')=\sqrt{\beta'^2-K_B^2}$,
with $\displaystyle K_B^2=\frac{2m}{\hbar^2} (2 \mu B)$. 
There are either two solutions or just one  depending on 
whether the value of $K_B$ (the onset of the hyperbola on the $\beta'$ axis)
is smaller or larger than $\beta_c'$. In the figure we have plotted the solutions for two values that exemplify both cases.
If  
$\displaystyle K_B \leq \beta_c'$
there are two solutions, whereas if 
$K_B> \beta_c'$ there is only one. Recall that in units of
inverse length $\displaystyle \left[(\alpha_1+\alpha_2)/2\right]$,
in this plot $\beta'_c=0.91$.}
\end{figure}


The full spectrum covers possibilities other than the one we have analysed.
Both channels can be open, in which case the effective coupling in
is no longer real; it is not difficult to repeat the 
calculations that led to (\ref{alpha.eq}) and find that it is
the continuation of the function
in (\ref{alpha.eq}) to complex arguments,
$\displaystyle \alpha_{{\rm eq}}(-i K')$, what appears here,
in the situation of boundary conditions of a plane wave 
entering the $u$-channel. The physical interpretation is clear and reflects
the fact that a fraction of the probability at the entrance leaks through
the other channel, which is also open.
As can be easily checked, the probability currents $J_u$ and $J_v$ are 
not separately conserved in the stationary state, but so is the sum 
of the two $J_u+J_v$ that involves the two components at once,  
\begin{equation}
\frac{d}{dz}\left(J_u+J_v \right)=\frac{\hbar}{2i m}
\frac{d}{dz}\left(u^*(z) \frac{du(z)}{dz}- u(z) \frac{du^*(z)}{dz}+
v^*(z) \frac{dv(z)}{dz}- v(z) \frac{dv^*(z)}{dz}\right)=0,
\label{current}
\end{equation}
which expresses the conservation of the total probability current.

In our case of open $u$-channel and closed $v$-channel, 
the wavefunction $v(z)$ in (\ref{sol1})
is real and its corresponding probability current $J_v=0$ vanishes, which
prevents any probability leakage. It is the reason
why $\displaystyle \alpha_{{\rm eff}}(\beta')$ is real
and why the probability conservation
can be cast in terms of the $J_u$ alone.

\vspace*{0.5cm}

Finally, let us briefly mention the possibility of discrete
bound states for a given value of the magnetic field
$B$, i.e., states with both 
$E \pm \mu B < 0$ negative. They are the solutions of,
\begin{equation}
\alpha_{{\rm eff}}(\beta')=\beta(\beta')\equiv\sqrt{\beta'^2-K_B^2},
\label{bound2}
\end{equation}
with $\beta=-iK$ and $\beta'=-iK'$ in (\ref{Ks}), and
$\displaystyle K_B^2=\frac{2m}{\hbar^2} (2 \mu B)$.
With $\alpha_1 \neq \alpha_2$, the number of solutions is either two or
just one, depending on  whether the value of $K_B$ is smaller or larger 
than $\beta_c'$, respectively (see Fig. 2).
In the limit of  
$K_B \gg \beta_c'$, the intesect of the hyperbola and the curve 
$\alpha_{{\rm eff}}(\beta')$ takes place for $\beta' > K_B$, 
in the region where
$\displaystyle \alpha_{{\rm eff}}(\beta'>K_B) \to 
\frac{\alpha_1 +\alpha_2}{2}$ is
asymptotically flat and the solution becomes,
\begin{equation}
\displaystyle E \approx -\mu B -
\frac{\hbar^2}{2m} 
\left( \frac{\alpha_1+\alpha_2}{2}\right)^2.
\label{bound}
\end{equation}

\section{Summary}
We have presented a simple example in one dimension which consists 
of a spin 1/2 particle submitted to a delta-type potential that interatcs 
with different 
strength with the spin up and spin down components of the wavefuntion,
and with an external magnetic field in the $x$-direction. Two coupled
channels for scattering are available. We have checked that in the case of
one open and one closed channels,
with a suitable
choice of the magnetic field a Feshbach resonance is produced so that in the 
neighbourhood of it the effective coupling flips sign, from
attractive to repulsive. The reflection ceofficient is enormalously
enhanced around it for all values of the incident wavenumber $K$,
and the scattering phase shift is $\delta_0 = \pi/2$.

\section{Appendix}
All the solutions presented in this article can be easily checked.
We have essentially used the following two facts \cite{Gott}, namely,
\begin{equation}
\left( \frac{d^2}{dz^2}+ k^2 \right) \; \frac{e^{ik|z|}}{2ik}=
\; \delta(z), \;\;\;
\left( \frac{d^2}{dz^2}-\beta^2\right) \; \frac{e^{-\beta|z|}}{2\beta}=
-  \delta(z).
\label{green}
\end{equation}

\vspace*{0.5cm}
Let us briefly review the spectrum of a one-dimensional
hamiltonian with an atractive delta-type potential:
\begin{equation}
-\frac{\hbar^2}{2m} \frac{d^2\chi(z)}{dz^2}-g \delta(z) \chi(z)=E \chi(z),
\label{1delta}
\end{equation}
For $E>0$ it becomes,
\begin{equation}
\left( \frac{d^2}{dz^2} + k^2 \right) \chi(z)=-2\alpha \chi(0) \delta(z),
\label{helm}
\end{equation}
where $\displaystyle \alpha=\frac{mg}{\hbar^2}$, and 
$\displaystyle k=\sqrt{\frac{2mE}{\hbar^2}}$.
With scattering boundary condition of an entering plane wave plus 
an outgoing scattered wave, the solution, according to (\ref{green})
is of the form,
\begin{equation}
\chi(z)= e^{ikz} - 2 \alpha \chi(0) \; \frac{e^{ik|z|}}{2ik},
\label{ls2}
\end{equation}
where, self-consistently, one finds, 
$\displaystyle \chi(0)=-\frac{1}{\displaystyle 1+\frac{\alpha}{ik}}$,
i.e.,
\begin{equation}
\chi(z)= e^{ikz} - \frac{\alpha}{ik+\alpha} e^{ik|z|}.
\label{ls}
\end{equation}
The reflection and transmission
amplitudes $r(k)$, $t(k)$
defined as,
$$\chi(z \to - \infty)\sim e^{ikz} + r(k) e^{-ikz},$$
\begin{equation}
\chi (z \to +\infty) \sim t(k) e^{ikz},
\label{rt}
\end{equation}
can be read off immediately,
\begin{equation}
r(k)=-\frac{\alpha}{ik+\alpha}, \;\;\; t(k)=\frac{ik}{ik+\alpha}.
\label{rt2}
\end{equation}

If $E<0$ 
this potential always holds only one bound state. Eq. (\ref{1delta})
becomes, with $\displaystyle \beta=\sqrt{\frac{2m|E|}{\hbar^2}}$,
\begin{equation}
\left( \frac{d^2}{dz^2} - \beta^2 \right) \chi(z)=-2\alpha \chi(0) \delta(z),
\label{helm2}
\end{equation}
and according to (\ref{green}),
\begin{equation}
\chi(z)=\frac{\alpha}{\beta} \chi(0) \;e^{-\beta |z|}.
\label{helm3}
\end{equation}
Setting $z=0$ one is led to conclude that 
\begin{equation}
\beta=\alpha,
\label{quantization}
\end{equation}
which is the quantization condition for the bound state energy
$\displaystyle E=-\frac{\hbar^2}{2m}\alpha^2$, whereas $\chi(0)$
remains free for normalization of the wave function:
\begin{equation}
\chi(z)=\sqrt{\alpha} \; e^{-\alpha |z|}.
\label{wf}
\end{equation}
The wave function of the bound state extends over a distance
$1/\alpha$ around $z=0$. 

\section{Acknowledgments}
We thank Prof. Nuria Barber\'an for reading the manuscript and for
helpful discussions and comments. We acknowledge financial support by
the Spanish Government through
the Consolider CPAN project CSD2007-00042, 
and by the Generalitat de Catalunya Program under contract 
number 2009SGR502.


\end{document}